\documentclass[aip,apl,graphicx]{revtex4-1} 

\usepackage{graphicx}
\usepackage{textcomp}
\usepackage{hyperref}

\draft 

\begin{document}

\title{GaAs droplet quantum dots with nanometer-thin capping layer for plasmonic applications} 

\author{Suk In Park}
\affiliation{Center for Opto-Electronic Materials and Devices Research, Korea Institute of Science and Technology, Seoul 136-791, South Korea}

\affiliation{Department of Physics and Astronomy, Seoul National University, Seoul 08-826, South Korea}

\author{Oliver Joe Trojak}
\affiliation{Department of Physics and Astronomy, University of Southampton, Southampton, SO17 1BJ, United Kingdom}

\author{Eunhye Lee}
\affiliation{Center for Opto-Electronic Materials and Devices Research, Korea Institute of Science and Technology, Seoul 136-791, South Korea}

\author{Jin Dong Song}
\email{jdsong@kist.re.kr}
\affiliation{Center for Opto-Electronic Materials and Devices Research, Korea Institute of Science and Technology, Seoul 136-791, South Korea}

\author{Jihoon Kyhm}
\affiliation{Center for Opto-Electronic Materials and Devices Research, Korea Institute of Science and Technology, Seoul 136-791, South Korea}

\author{Ilki Han}
\affiliation{Center for Opto-Electronic Materials and Devices Research, Korea Institute of Science and Technology, Seoul 136-791, South Korea}

\author{Jongsu Kim}
\affiliation{Department of Physics, Yeungnam University, Gyeongsangbuk-Do 712-749, Korea}

\author{Gyu-Chul Yi}
\affiliation{Department of Physics and Astronomy, Seoul National University, Seoul 08-826, South Korea}

\author{Luca Sapienza}
\email{l.sapienza@soton.ac.uk}\homepage{www.quantum.soton.ac.uk}
\affiliation{Department of Physics and Astronomy, University of Southampton, Southampton, SO17 1BJ, United Kingdom}

\date{\today}

\begin{abstract}

We report on the growth and optical characterisation of droplet GaAs quantum dots with extremely-thin (11\,nm) capping layers. To achieve such result, an internal thermal heating step is introduced during the growth and its role in the morphological properties of the quantum dots obtained is investigated via scanning electron and atomic force microscopy.  Photoluminescence measurements at cryogenic temperatures show optically stable, sharp and bright emission from single quantum dots, at near-infrared wavelengths. Given the quality of their optical properties and the proximity to the surface, such emitters are ideal candidates for the investigation of near field effects, like the coupling to plasmonic modes, in order to strongly control the directionality of the emission and/or the spontaneous emission rate, crucial parameters for quantum photonic applications.

\end{abstract}


\maketitle 

Single epitaxial quantum dots (QDs) are bright and stable sources of quantum light that have been implemented in a wide range of quantum optics and quantum technology experiments \cite{Gazzano, Hofling}. To preserve the quality of the optical properties of quantum dots grown by self-assembly Stranski-Krastanow techniques, these emitters are typically buried under a capping layer exceeding 50\,nm of thickness \cite{capping}. In order to control their emission properties, QDs can be coupled to confined or propagating modes, like those from optical cavities or waveguides, in cavity quantum electrodynamics experiments \cite{QED}. Another very powerful approach for the control of the emission of solid-state light sources relies on their coupling to confined plasmonic modes \cite{plasmon1}. However, such plasmonic modes are effective at short (nanometer-scale) distances, therefore emitters, like fluorescent molecules or colloidal quantum dots, are generally deposited on the devices via spin coating or drop casting, to ensure that the emitter is in close proximity to the localised electromagnetic field \cite{quantum_plasmonics}. Such emitters, however, often suffer from limited optical stability, thus reducing the time-scale and range of experiments that can be performed with them. In order to couple epitaxial single quantum dots to near-field confined modes, the emitter needs to be in close (few nanometers range) proximity to the metal-dielectric interface where the plasmonic mode exists \cite{plasmon2} and this has, thus far, made them unsuitable for plasmonic experiments, given the presence of a thick ($>$\,50\,nm) capping layer, used to ensure optimal optical properties. Attempts to observe plasmonics effects with quantum dots have been made using droplet quantum dots \cite{droplet} with 50\,nm capping layer in cathodoluminescence but the non-optimal optical quality and thick capping layer did not allow for the plasmonic coupling to be observable optically \cite{droplet_plasmonics}.
We have grown and optically characterised GaAs droplet quantum dots with a very thin (11\,nm) capping layer. By investigating the structural and optical properties of these emitters, we show that the technique that we have developed allows the growth of stable emitters with good optical quality that are ideal candidates for investigating plasmonic and near-field effects in general. 

The size, shape and density are crucial parameters for the control over the optical characteristics of QDs: the control of these factors is critical in the growth of QDs for quantum photonics applications. 
Droplet epitaxy has been used to grow low density QDs \cite{droplets, droplets2}, however, since this method is based on the separation of group III and V elements, a post-annealing process, such as a rapid thermal annealing, is generally conducted in order to improve the crystallinity of the droplet QDs \cite{RTA}. By growing QDs at a high substrate temperature, their optical properties are improved, however, the shape of QDs could be deformed and QDs can even disappear at high substrate temperature due to heating. In order to overcome this drawback, it was shown that a thin capping layer can be deposited to avoid the deformation of QDs during high-temperature growth \cite{high_T}. 
We propose an internal thermal heating (ITH) process that, compared to rapid thermal annealing, has the merit of annealing in the ultra-high vacuum chamber. The ITH also allows a better control over the size and shape of the grown QDs. In addition, the capping layers can be grown at the ITH temperature, that is higher compared to the temperature at which GaAs droplet QD are formed. ITH also allows to control the height of QDs, a key factor for carrier quantum confinement. Furthermore, given the strain-free nature of GaAs droplet QDs on AlGaAs, this procedure allows the growth of very thin (few nanometer) capping layers. 
	
In this study, we investigate the growth of GaAs/Al$_{0.3}$Ga$_{0.7}$As droplet epitaxial QDs grown with an ITH process. The size and density of ITH GaAs QDs is measured via atomic force microscopy (AFM) and the optical properties are assessed via micro-photoluminescence.

We grow GaAs droplet epitaxial QDs on Al$_{0.3}$Ga$_{0.7}$As/GaAs substrates and use an ITH process to modify the morphology of the QDs and enhance their optical properties. The growth procedure is shown in Fig.\,1a. After deoxidation at $\sim$600$^{\circ}$C under As$_{4}$, a 50\,nm-thick Al$_{0.3}$3Ga$_{0.7}$As layer is grown on a $\sim$200\,nm-thick GaAs buffer layer. On this layer, surface Ga droplets, GaAs droplet islands with no ITH, and GaAs islands including an ITH step are grown for comparison. The growth conditions of Ga droplets are: substrate temperature of 321$^{\circ}$C, 2 Ga monolayer (ML) coverage, and 0.5 ML/s in Ga flux equivalent to the growth rate of GaAs. In order to achieve uniform growth, we used a growth interruption time of 10\,s after growing Ga droplets.
An As$_{4}$ flux of 1$\times$10$^{-5}$ Torr is injected on the liquid Ga droplets, near room temperature, after cooling down and the substrate temperature for ITH (T$_{ITH}$) is varied in the range 300-580$^{\circ}$C. The ITH needed for the formation of GaAs islands is conducted for 10 minutes, at an As$_{4}$ flux of 6$\times$10$^{-6}$ Torr. As previously reported \cite{blueshift}, the ITH step can blue-shift the emission wavelength of the quantum dots, due to a modification of the shape and size of the emitter. A thin, 11\,nm-thick, Al$_{0.3}$Ga$_{0.7}$As capping layer is grown on the GaAs islands after the ITH process: at first, an 8\,nm-thick Al$_{0.3}$Ga$_{0.7}$As capping layer is deposited on the ITH GaAs islands at 563$^{\circ}$C, then, the additional capping layer of 3 nm thickness is grown at 580$^{\circ}$C. Typically, an initial capping layer is deposited at a low substrate temperature to avoid deformation of the QDs. However, the low-temperature growth of the capping layer can reduce the optical quality of the emitters. Since in our case the capping layer is deposited at an ITH temperature of 563$^{\circ}$C, this can reduce the formation of defects and decrease the emission linewidth of the confined excitons \cite{capping2}: it is therefore a route for improving the optical qualities of low density GaAs droplet epitaxial QDs for quantum photonics applications.

By investigating, via AFM, the QDs morphology as a function of the ITH temperature, we see that the height of GaAs QDs can be controlled by the ITH process with no significant change in the QD width and density (see Fig.1b,c,d). The height of the GaAs islands starts decreasing when the ITH temperature reaches about 500$^{\circ}$C. At a temperature of 563$^{\circ}$C, the height of the GaAs islands is measured to be 8.7$\pm$3.4\,nm and the width 104.2$\pm$13.4\,nm. The GaAs QDs obtained at this temperature are then capped with an 11\,nm-thick AlGaAs layer (see Fig.\,1b).
As shown in Fig.\,2b, the width is practically unchanged between the Ga droplets (labeled as "Metal"), and the GaAs islands (labeled as "No heating"). The height of the GaAs islands is about two times higher than the height of Ga droplets and tends to decrease at temperatures T$_{ITH}$ exceeding 500$^{\circ}$C.
Such trends are clearly shown when looking at the aspect ratios (height/width) in Fig.\,1c: the ITH process for GaAs islands mainly changes the height but not the width of GaAs islands, with a flattening of the top of the islands after annealing, consistent with previous reports \cite{flattening}. The elongated quantum dot shape is attributed to the different diffusivity (along [110] and [1-10]) of Ga on GaAs(001) \cite{elongated}.

The density of Ga droplets, GaAs islands, and ITH GaAs islands is shown in Fig.\,1d and appears to be the same for the three different species under consideration and is of about 6 droplets or islands per micrometer square (see also Fig.\,2c). Concerning droplet epitaxial growth, the change in density of GaAs islands after crystallization at a low substrate temperature is typically negligible. In the case of Ga droplets, it significantly changes as a function of the substrate temperatures, a phenomenon which can be explain by Ostwald ripening, whereby larger clusters grow at the expenses of small clusters, thus decreasing the total density. However, no significant density change of the ITH GaAs islands is observed, although we note that the effect of Ostwald ripening can be accelerated by an increase of the substrate temperature \cite{Ostwald}. The present result implies that GaAs droplet islands can endure a T$_{ITH}$ of $\sim$500$^{\circ}$C with no increment of droplet size at the expenses of smaller islands but with a reduction of material from the top of the islands.  
	
During our growth, the ITH process is conducted in an As$_{4}$ atmosphere of 6$\times$10$^{-6}$ Torr to prevent severe deformation of GaAs islands due to re-evaporation of As$_{4}$ by heating. The results in Figures 1b,c,d show that the height of the GaAs islands can be controlled by ITH without any significant modification to their width and density. Given the control over the height of the ITH GaAs islands, deposition of even thinner capping layers, while still preserving quantum confinement effects, is expected to be possible.

In Fig.\,2b, we show that, given the proximity of the GaAs islands to the surface, they are visible under a scanning electron microscope (SEM), therefore allowing an easy deterministic integration within nanophotonic devices, using SEM techniques developed for buried InAs/GaAs QDs \cite{SEM_PhC}. The shape of the QDs is elliptically elongated along the (1 -1 0) direction, with height and (long axis) width of the surface features of 3.8$\pm$1.0\,nm and 208.3$\pm$42.5\,nm, respectively. 
Fig.\,2c and 2d show AFM images of the GaAs islands with 11\,nm-AlGaAs capping layer, carried out in tapping mode, using a tetrahedral, point-terminated, silicon cantilever, with a tip radius of 7\,nm, spring constant of 26\,N/m and resonance frequency of 300\,kHz (nominal values). We measure a density of about 6 QDs/$\mu$m$^2$ and, considering that the average height of the QDs before capping is of 8.7$\pm$3.4\,nm (and width of 104.2$\pm$13.4\,nm), the actual distance between the QD and the surface is of just a few nanometers (see Fig.\,2a). 

We then characterise the emission properties of the droplet QDs, by means of photoluminescence imaging \cite{NComm} and micro-photoluminescence spectroscopy (for more details about the setup, see Ref.\cite{And}). By illuminating the sample with a 455\,nm light emitting diode, at cryogenic temperatures, we assess the wide-field photoluminescence, as shown in Fig.\,3a: several single emitters are visible, proving that most of the QDs are optically active. We then select single emitters and excite them with a 405\,nm continuous-wave laser, with an excitation spot size of about 2\,$\mu$m in diameter. The emitted light is sent to a grating spectrometer and an example of the micro-photoluminescence spectrum collected by a silicon charge-couped device is shown in Fig.\,3b: sharp emission lines (with linewidths down to 0.175\,nm, as shown in the inset of Fig.\,3b) are visible at wavelengths around 730\,nm. These results show that the GaAs islands with ultra-thin capping layer allow quantum confinement of the carriers and stable emission in the near-infrared range of wavelengths. The reduction of the capping layer thickness implies that the emitters are closer to surface states that can cause a broadening of the emission lines \cite{broadening}. We note that linewidths one order of magnitude lower have been reported for droplet quantum dots buried under a 100\,nm-thick capping layer \cite{linewidth}. However, because of the larger distance between the emitter and the surface, such emitters, would be unsuitable for surface near-field applications. 
By varying the excitation power, we can assess the brightness of the sources, as shown in Fig.\,3c where the emission intensity (defined as photon flux on the first lens, thus taking into account the transmission of the setup that is measured to be $\sim$9.2\,$\%$) is plotted as a function of laser pump power. Such results are reproducible over different excitation runs and cool downs, proving the stability of the light emission from these droplet QDs, even in the presence of a very thin capping layer.

Such emitters are therefore ideal for studying near-field plasmonic effects, for instance by depositing metallic structures around them, like metallic rings \cite{rings} that have been developed to increase the brightness of single-photon emission, bowtie antennas \cite{bowties} or metallic cubes \cite{cubes} implemented to enhance the light-matter interaction. Compared to fluorescent emitters \cite{plasmonic_emitters}, like $J$-aggregates, dye molecules, colloidal quantum dots that have mostly been used in plasmonic experiments so far, the droplet QDs here developed grant access to sharp emission lines and stable emission properties that, thanks to the capping layer, are not degraded after laser illumination, interaction with the environment or subsequent cool downs. Therefore, we expect that such emitters will enable the investigation of near-field effects and possible coherent energy transfer between single emitters and confined or propagating near-field optical modes.
Another advantage of having a thin capping layer lies in the fact that the QD position can be retrieved in a much easier way compared to buried QDs (as shown in Fig.\,2): techniques like combined photoluminescence and atomic force microscopy \cite{SR} could thus be easily and effectively implemented to study growth characteristics and/or deterministically embed single emitters within engineered photonic devices.

\begin{acknowledgments}
We would like to thank Kartik Srinivasan for giving us access to the atomic force microscope at the Center for Nanoscale Science and Technology at the National Institute of Standards and Technology.
The authors from KIST acknowledge support from the KIST institutional flagship program and the GRL Program. L.S. acknowledges financial support from EPSRC, grant EP/P001343/1.
\end{acknowledgments}

\newpage

\begin{figure}
 	\includegraphics[width=\columnwidth]{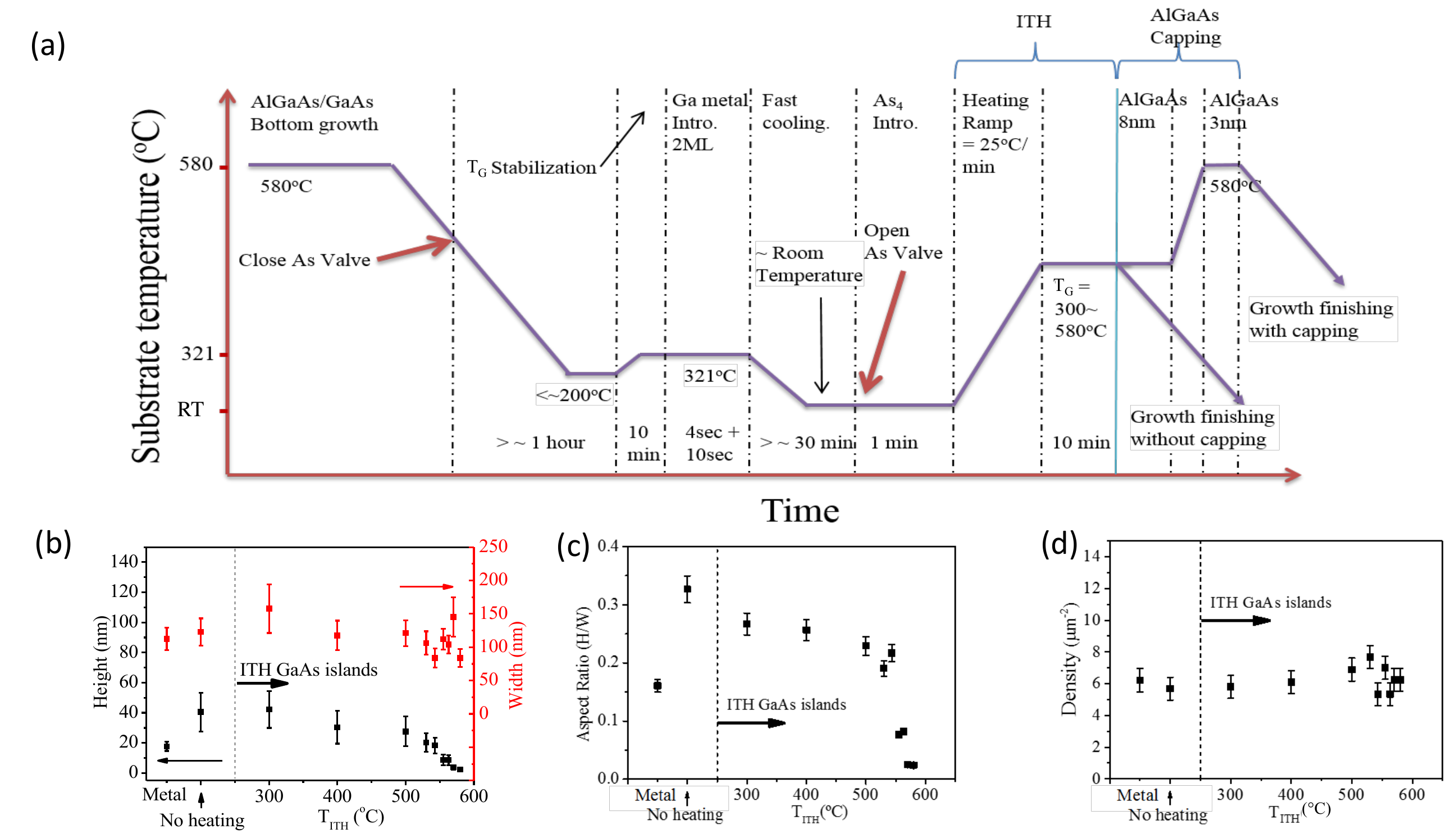}
 	\caption{(a) GaAs droplet QDs growth procedure with an internal thermal heating (ITH) step. (b) Dimensions (width and height, black and red symbols respectively), (c) aspect ratio (height/width, H/W) and (d) density of the QDs with no ITH and as a function of ITH temperature (T$_{ITH}$). (T$_{G}$\,=\,growth temperature, UHV\,=\,ultra high vacuum, RT\,=\,room temperature, ML\,=\,monolayer).}
 \end{figure}

\begin{figure}
 	\includegraphics[width=\columnwidth]{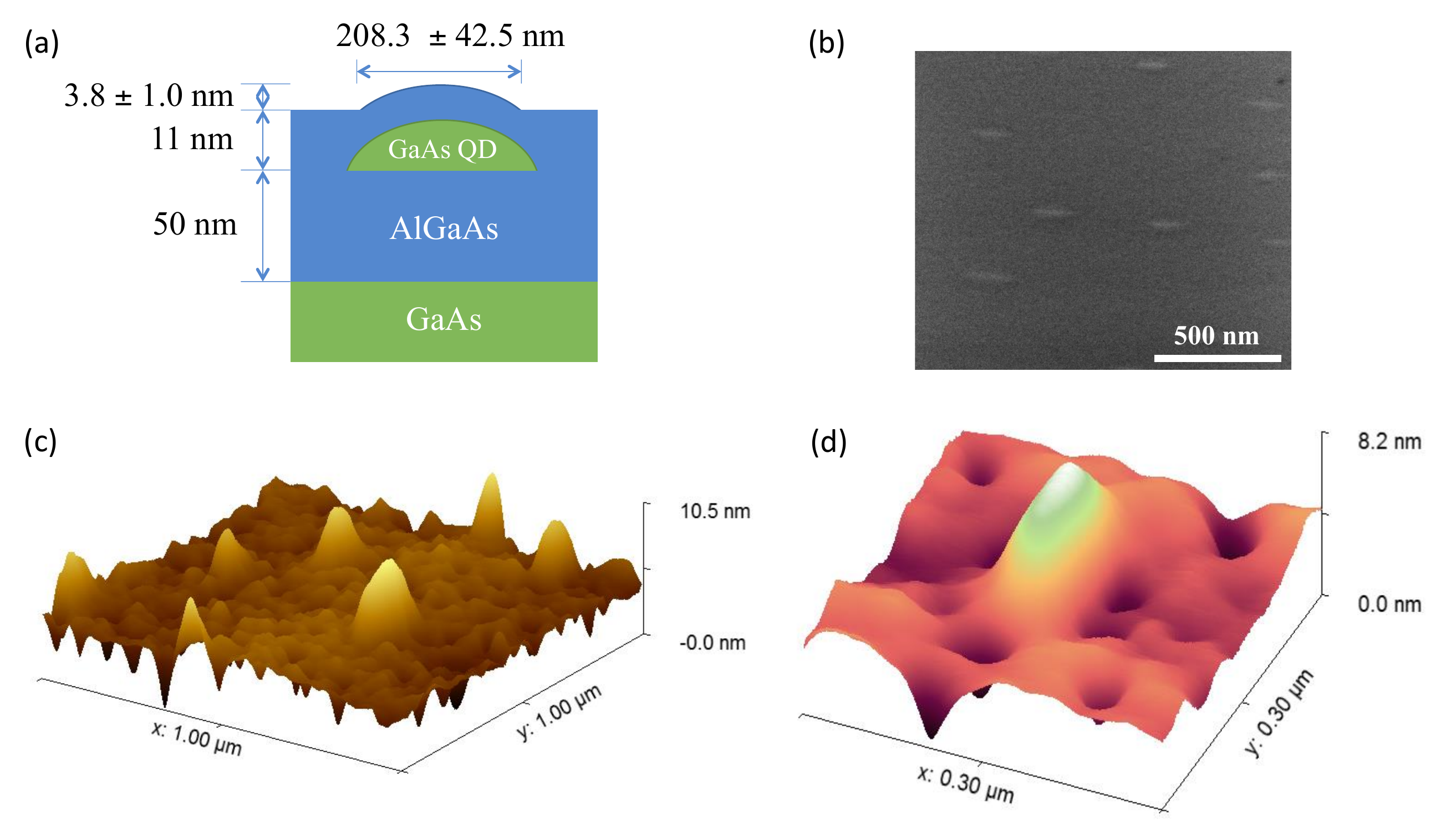}
 	\caption{(a) Schematic (not to scale) of the droplet QDs, showing growth thicknesses and average dimensions, as measured from several capped QDs. (b) Scanning electron micrograph showing droplet QDs. (c) Wide-range atomic force micrograph showing a density of about 6 QDs/$\mu$m$^2$. (d) Atomic force micrograph of a single droplet QD.}
 \end{figure}

\begin{figure}
 	\includegraphics[width=0.7\columnwidth, angle=-90]{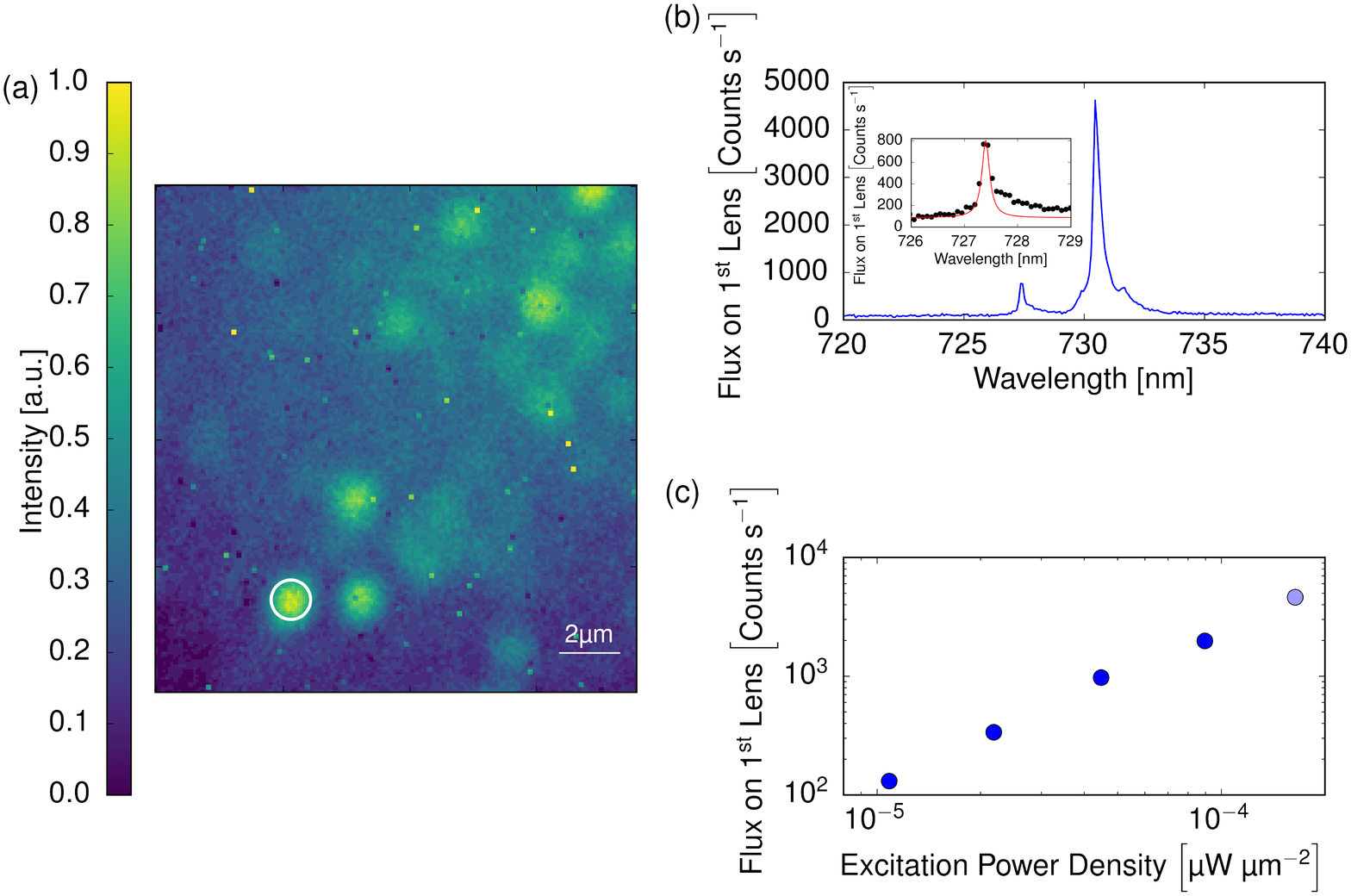}
 	\caption{(a) Photoluminescence image showing the emission of several QDs, collected under 455\,nm light emitting diode illumination by an electron-multiplied charge-coupled device (EMCCD). (b) Photoluminescence spectrum (from the quantum dot highlighted by the circle in panel (a)) collected under 405\,nm continuous-wave laser excitation with power density of 1.8$\times$10$^{-4}$ $\mu$W/$\mu$m$^{2}$, at a temperature of 4.7\,K, by a CCD camera at the exit of a grating spectrometer. Inset: Zoom-in of the photoluminescence emission line at 727.4\,nm from the spectrum in the main panel (symbols) and its Lorentzian fit (solid line), showing a linewidth of 0.175$\pm$0.022\,nm. (c) Emission intensity plotted as a function of laser excitation power for the emission line at 730.5\,nm shown in panel (b). The lighter blue symbol represents data obtained under the same conditions as panel (b).}
 \end{figure}
 

\begin{thebibliography}{50}


\bibitem{Gazzano} O. Gazzano, G.S. Solomon, J. Opt. Soc. Am. B  \textbf{33}, C160 (2016).

\bibitem{Hofling} C.P. Dietrich, A. Fiore, M.G. Thompson, M. Kamp, S. H\"ofling, Laser Photonics Review \textbf{10}, 870 (2016).

\bibitem{capping} C.F. Wang, A. Badolato, I. Wilson-Rae, P.M. Petroff, E. Hu, Appl. Phys. Lett. \textbf{85}, 3423 (2004).

\bibitem{QED} M. Pelton, Nature Photonics \textbf{9}, 427 (2015).

\bibitem{plasmon1} C. Vandenbem, D. Brayer, L.S. Froufe-Perez, and R. Carminati, Phys. Rev. B \textbf{81}, 085444 (2010).

\bibitem{quantum_plasmonics} F. Marquier, C. Sauvan, J.-J. Greffet, ACS Photonics \textbf{4}, 2091 (2017).

\bibitem{plasmon2} S.A. Maier, \textit{Plasmonics: Fundamentals and Applications}, Springer (2007).

\bibitem{droplet} N. Koguchi and K. Ishige, Jpn. J. Appl. Phys. \textbf{32}, 2052 (1993).

\bibitem{droplet_plasmonics} G. Nouges, Q. Merotto, G. Bachelier, E.H. Lee, J.D. Song, Appl. Phys. Lett. \textbf{102}, 231112 (2013).

\bibitem{droplets} T. Mano, T. Kuroda, S. Sanguinetti, T. Ochiai, T. Tateno, J. S. Kim, T. Noda, M. Kawabe, K. Sakoda, G. Kido, and N. Koguchi, Nano Lett. \textbf{5}, 425 (2005).

\bibitem{droplets2} S.-K. Ha, J.D. Song, S.Y. Kim, J.I. Lee, S. Bounouar, L.S. Dang, J.S. Kim, Journal of the Korean Physical Society \textbf{58}, 1330 (2011).

\bibitem{RTA} P. Moon, J.D. Lee, S.K. Ha, E.H. Lee, W.J. Choi, J.D. Song, J.S. Kim, L.S. Dang, Phys. Status Solidi RRL \textbf{6}, 445 (2012).

\bibitem{high_T} M. Jo, T. Mano, and K. Sakoda, J. Appl. Phys. \textbf{108}, 083505 (2010).

\bibitem{blueshift} S. Bietti, J. Bocquel, S. Adorno, T. Mano, J.G. Keizer, P.M. Koenraad, S. Sanguinetti, Phys. Rev. B \textbf{92} 075425 (2015).

\bibitem{capping2} K. Kuroda, T. Kuroda, K. Watanabe, T. Mano, G. Kido, N. Koguchi, and K. Sakoda, J. Lumin. \textbf{130}, 2390 (2010).

\bibitem{flattening} M. Jo, T. Mano, and K. Sakoda, J. Appl. Phys. \textbf{108}, 083505 (2010).

\bibitem{elongated} C. Somaschini, S. Bietti, S. Sanguinetti, N. Koguchi, A. Fedorov, Nanoscale Research Letters  \textbf{5}, 1865 (2010).

\bibitem{Ostwald} C. Heyn, A. Stemmann, A. Schramm, H. Welsch, W. Hansen, and A. Nemcsics, Phys. Rev. B \textbf{76}, 075317 (2007).

\bibitem{SEM_PhC} K. Kuruma, Y. Ota, M. Kakuda, D. Takamiya, S. Iwamoto, and Y. Arakawa, Appl. Phys. Lett. \textbf{109}, 071110 (2016).

\bibitem{NComm} L. Sapienza, M. Davanco, A. Badolato and K. Srinivasan, Nat. Commun. \textbf{6}, 7833 (2015).

\bibitem{And} T. Crane, O.J. Trojak, J.P. Vasco, S. Hughes, L. Sapienza, ACS Photonics \textbf{4}, 2274 (2017).

\bibitem{broadening} N. Ha, T. Mano, Y.-L. Chou, Y.-N. Wu, S.-J. Cheng, J. Bocquel, P.M. Koenraad, A. Ohtake, Y. Sakuma, K. Sakoda, T. Kuroda, Phys. Rev. B \textbf{92}, 075306 (2015).

\bibitem{linewidth} T. Mano, M. Abbarchi, T. Kuroda, C.A. Mastrandrea, A. Vinattieri, S. Sanguinetti, K. Sakoda, M. Gurio, Nanotechnology \textbf{20}, 395601 (2009).

\bibitem{rings} O.J. Trojak, S.I. Park, J.D. Song, L. Sapienza, Appl. Phys. Lett. \textbf{111} 021109 (2017).

\bibitem{bowties} H. Fischer and O.J.F. Martin, Optics Express \textbf{16}, 9144 (2008).

\bibitem{cubes} L.J. Sherry, S.-H. Chang, G.C. Schatz, R.P. Van Duyne, Nano Lett. \textbf{5}, 2034 (2005).

\bibitem{plasmonic_emitters} P. T\"orm\"a and W.L. Barnes, Rep. Prog. Phys. \textbf{78}, 013901 (2015).

\bibitem{SR} L. Sapienza, J. Liu, J.D. Song, S. Falt, W. Wegscheider, A. Badolato, K. Srinivasan, Scientific Reports \textbf{7}, 6205 (2017).



\end{thebibliography}
\end{document}